\documentclass[11pt]{article}
\usepackage[a4paper, top=1.8cm, bottom=1.8cm, left=2.0cm, right=2.0cm]{geometry}
\usepackage{multicol}
\usepackage{setspace}
\usepackage{wrapfig}

\usepackage[T1]{fontenc}
\usepackage[utf8]{inputenc}
\usepackage{helvet}         % This is the pdfLaTeX equivalent to Arial
\usepackage{graphicx}
\usepackage{amssymb}
\usepackage{epstopdf}
\usepackage{wrapfig}
\usepackage{color}
\usepackage{chemformula}
\usepackage{caption}
\usepackage{upgreek}
\usepackage[dvipsnames]{xcolor}
\usepackage[font={color=black},figurename=Fig.,labelfont={it}]{caption}
\usepackage{siunitx}

\usepackage{hyperref}
\usepackage{xcolor}
\usepackage{enumitem} 
\usepackage{paralist}
\setitemize{noitemsep,topsep=0pt,parsep=0pt,partopsep=0pt,leftmargin=1.5\parindent}
\hypersetup{
colorlinks=true,
linkcolor=blue,
urlcolor=blue,
citecolor=blue,
}
\usepackage{natbib}
\usepackage{aasmacros}
\usepackage{doi}
\usepackage[labelformat=empty]{caption}
\setcitestyle{aysep={},yysep={}}

\usepackage{soul}

\newcommand{\squishlist}{
 \begin{list}{$\bullet$}
  { \setlength{\itemsep}{1pt}
     \setlength{\parsep}{0pt}
     \setlength{\topsep}{1pt}
     \setlength{\partopsep}{0pt}
     \setlength{\leftmargin}{1.5em}
     \setlength{\labelwidth}{1.5em}
     \setlength{\labelsep}{0.5em} } }
\newcommand{\squishend}{
  \end{list}  }

\begin{document} 

\title{Detecting habitable exoplanet atmospheres with LIFE, the Large Interferometer for Exoplanets. (Thematic area: Astro)}
\author{\parbox{\textwidth}{\centering Sarah Rugheimer$^1$, Aiden Weatherbee, James Fecanin, Alistair Glasse, Paul Rimmer, Eleonora Alei, Esther Wang, Marrick Braam, Sascha P. Quanz, Adrian Glauser, Alexander Archibald, Beth Biller, Mark Booth, Tereza Constantinou, Gregory Cooke, Daniel Dicken, Trent Dupuy, Mei Ting Mak, Paul Palmer, Tim Pearce, Vito Squicciarini, Amaury Triaud, Floris van der Tak, and Sergey Yurchenko. \emph{Co-signatories at end.} \\ 
$^1$\small{Lead author: Sarah Rugheimer, Institute for Astronomy, University of Edinburgh, Blackford Hill, EH9 3HJ
 \url{s.rugheimer@ed.ac.uk}}}}
\date{28 November 2025}

\maketitle

\vspace{-40pt}

\section{Executive Overview}

A key goal of astronomers with the next generation telescopes is to detect signs of life in exoplanet atmospheres. NASA's next flagship is the Habitable Worlds Observatory (HWO). In the context of ESA's Voyage 2050 program, the Senior Committee report prioritises detecting habitable exoplanet atmospheres in the mid-IR. The most suited mission for this is the Large Interferometer for Exoplanets (LIFE) which can detect an even wider range of biosignatures than HWO and at lower concentrations. LIFE is a global science collaboration based out of ETH Zürich. With the UK's expertise in building infrared instruments we could play a leading role in realising an ambitious European-led mission. Notably, LIFE is able to detect necessary planetary context like surface temperature and pressure, along with a key discriminator molecule for biosignature false positives, methane, which will be much harder or impossible with HWO. Also, LIFE will be able to investigate many of the nearby rocky exoplanets known from radial velocity searches that are inaccessible to HWO due to its limited spatial resolution. 

\vspace{-10pt}

\section{Scientific Motivation \& Objectives}

Surveys from Kepler and TESS have demonstrated that small planets ($\lesssim$4R$_\oplus$) are very common \citep[e.g.][]{dressing2013}, and that rocky planets in the habitable zone (HZ) occur around a substantial fraction of Sun-like and M-dwarf stars \citep[e.g.][]{Bryson2020}. With the James Webb Space Telescope (JWST) we entered a new era of exoplanet science. By providing precise transit and secondary eclipse spectroscopy, it has detected molecules in sub-Neptunes and gas giants (e.g. H$_2$O, CO$_2$, SO$_2$ in WASP-39b \citep{Rustamkulov2023}), laying the groundwork for retrieving mid-IR spectra. For temperate, rocky exoplanets JWST has shown clear limitations -- both for individual observations, but also in terms of number of objects that can be probed. For instance, planetary signals can be contaminated by the stellar signal \citep{rackham2019} and even in larger planets the detection of a molecules can be controversial \citep{madhusudhan2025, welbanks2025, niraula2025}, indicating that the next generation of space-based telescopes will be required for robust life detection.

\begin{figure}[h!]
\centering
    \includegraphics[width=1.0\textwidth ]{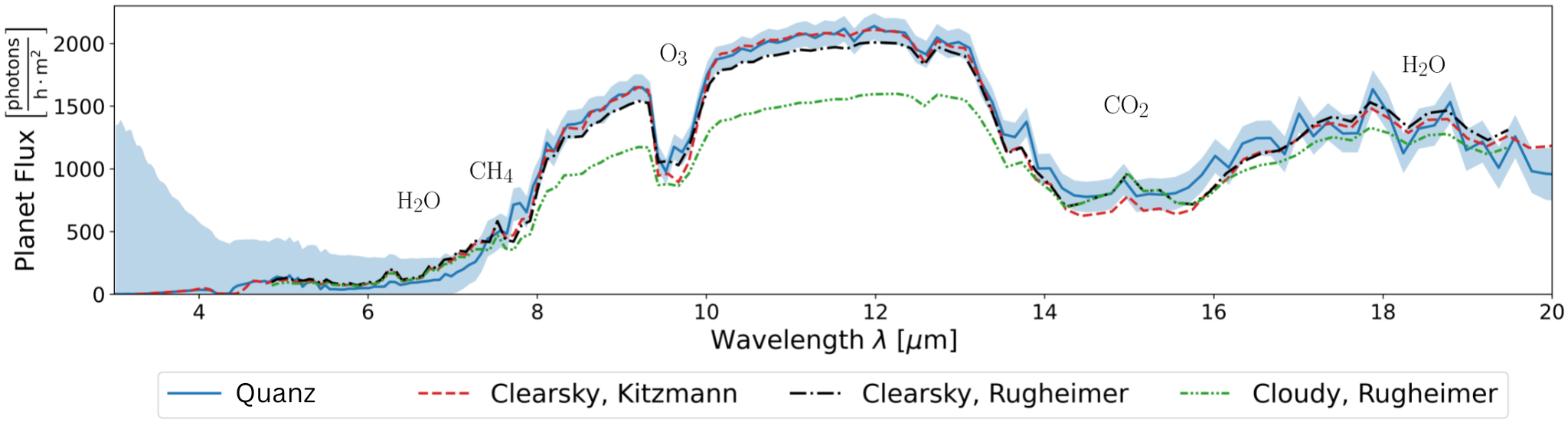} \caption{Fig. 1: Earth-twin at 10pc emission spectrum showing the dominant atmospheric features in the mid-IR for H$_2$O, CO$_2$, CH$_4$, and O$_3$. The blue-shaded region shows noise calculated from \texttt{LIFEsim}. Figure from \citep{konrad2022}.}  \label{spectrum}
\end{figure}

There is a promising path forward: a dedicated mid-infrared nulling interferometer such as the Large Interferometer For Exoplanets (LIFE) mission \citep{Quanz2022_LIFEI}. This is a large class mission concept that builds on the heritage of Europe’s Darwin and NASA’s Terrestrial Planet Finder – Interferometer (TPF-I). LIFE is designed to operate at the Sun–Earth L2 point, with four collector spacecraft equipped with $\sim$3-metre apertures and a fifth beam-combiner spacecraft. Its primary goal is to deliver direct-imaging thermal-emission spectra (requirement 6–16 µm, goal 4–18.5 µm, R = 100) that accurately constrain planetary radius, effective temperature, pressure-temperature profiles, and the abundances of key biosignature and habitability-relevant molecules, including  H$_2$O, CO$_2$, O$_3$, and CH$_4$ (see Fig. \ref{spectrum}).

\vspace{-10pt}

\subsection{Assessing Habitability} 

The mid-IR is a critical wavelength region for assessing planetary habitability as it provides a direct measurement of the planet's thermal emission and surface temperature. Surface temperature is not accessible in the VIS/NIR with HWO. 

%\begin{figure}[h!]
\begin{wrapfigure}{R}{0.6\textwidth}
\centering
\vspace{-25pt} % tighten space above
    \includegraphics[width=0.6\textwidth ]{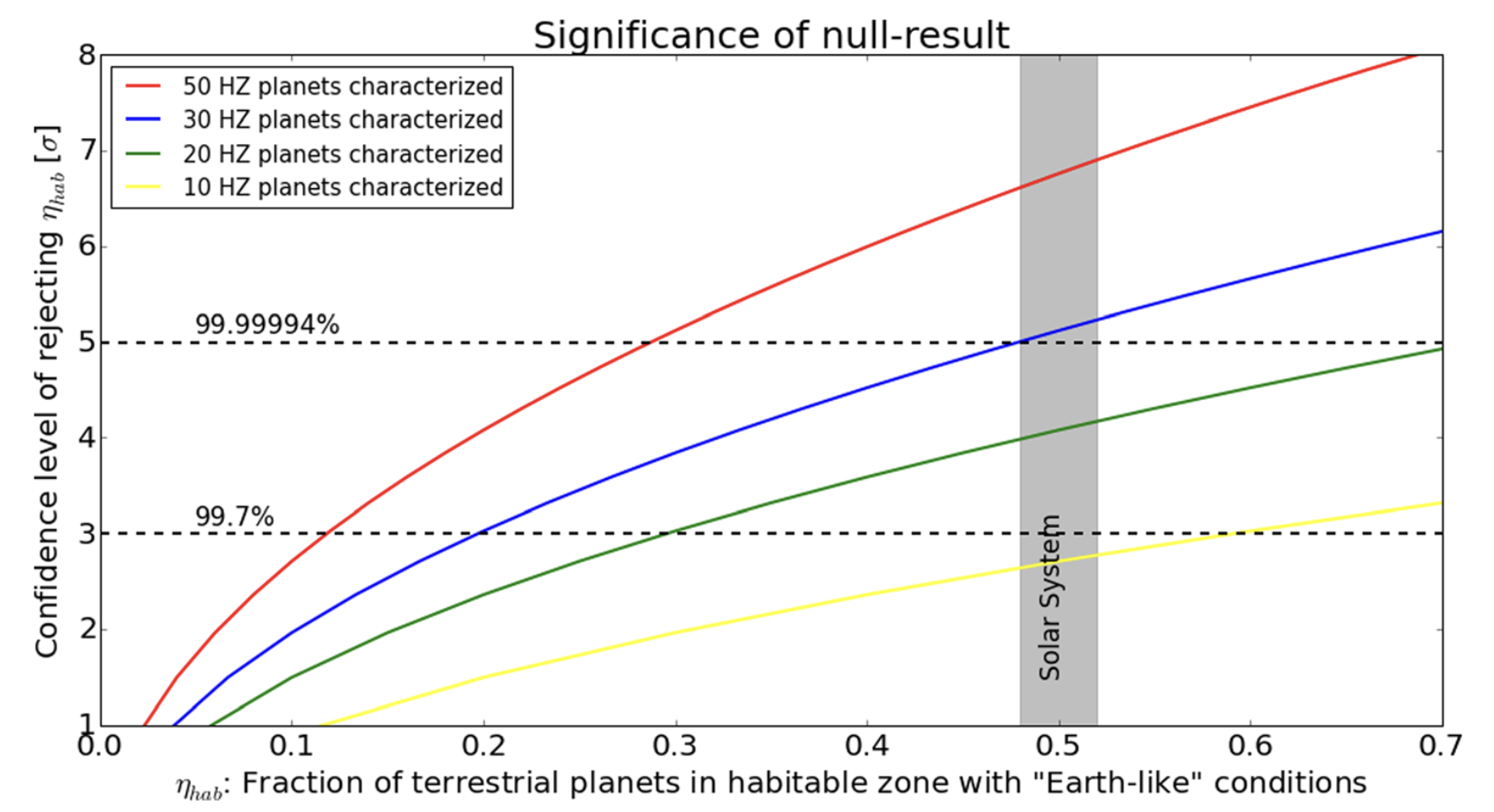} \caption{Fig. 2: The statistical significance of the null result for a given number of characterised exoplanets. Figure from \citep{quanz2022}.}  \label{null}
%\end{figure}
\vspace{-11pt} % tighten space above

\end{wrapfigure}

LIFE can sufficiently constrain the planet's radius, surface temperature, and water content to confirm the stability of surface oceans \citep{rugheimer2026} and habitability for life as we know it (and thus life as we can \emph{detect} it). Additionally, LIFE can constrain H$_2$O and CO$_2$, two key molecules to provide additional planetary context for habitability and measuring carbon depletion compared to other planets in a system may provide a secondary indicator of liquid water oceans and potentially biomass \citep{triaud2024}. 

%\subsubsection{Statistical significance and a potential null result}

No current mission will provide a large enough statistical sample of habitable exoplanets to assess any significance of the frequency of planets with Earth-like conditions. LIFE, however, is expected to characterise a statistically significant number of exoplanets to also shed light on the significance if we do not find signs of habitability (see Fig. \ref{null}). If LIFE characterises its goal of 50 habitable planets and finds no Earth-like conditions, then we can reject the hypothesis that 10\% of planets in the HZ are indeed habitable with a 3$\sigma$ confidence \citep{quanz2022}.

\vspace{-10pt}

\subsection{Biosignatures}

Research has shown that the most abundant and common biosignatures have notable false positives \citep[see][for a review]{schwieterman} and thus there is no single molecule that will be convincing for life. The strongest biosignature combination remains a chemical disequilibrium in the atmosphere. The classic example is the co-existence of an oxidising gas like O$_2$ or its proxy O$_3$, with a reducing gas such as CH$_4$ (e.g. \citep[see][]{lederberg1965,lovelock1965,sagan1993, totton2018}. Without a consistent biological source, these compounds destroy each other rapidly in an atmosphere. For robust biosignature identification we need an improved understanding of the underlying chemical network \citep{winiberg2025}.

LIFE is specially designed to measure this disequilibrium. Its 6-16 µm (goal 4-18.5 µm) bandpass captures the 7.7 µm CH$_4$ and 9.6 µm O$_3$ features which are detectable at R=100 \citep{konrad2024}. Ozone is a photochemical proxy of O$_2$, and CH$_4$ in particular is critical for distinguishing the known false positives of oxygen \citep{meadows2018}. Methane's strongest spectral absorption feature is in the mid-IR and it is inaccessible to HWO, especially for Earth-like concentrations \citep{kawashima2019}. Work has also shown that LIFE is able to detect biosignatures at a earlier Earth-like geological epochs before the rise of atmospheric O$_2$ \citep{alei2022} as well as detecting other potential biosignature gases such as PH$_3$, N$_2$O, CH$_3$Cl, and CH$_3$Br \citep{angerhausen2023, angerhausen2024}.

LIFE can also measure the abundances of major prebiosignature molecules such as HCN and HC$_3$N that signal the promise of life's origin on a planet \citep{Rimmer2021, Claringbold2023}, and thereby provide key context needed to assess biosignature claims \citep{Catling2018}. 

% Longer version from Paul for posterity - Understanding how life begins is essential for interpreting biosignatures. Without an origins-of-life framework, we risk repeating the same unproductive debates about whether a given spectral feature “could be abiotic.” Prebiosignatures, species that can form without biology but are linked to known prebiotic pathways, can provide the critical context needed to reframe this narrative. LIFE will be able to detect such prebiosignatures as HCN and HC3N, building on early demonstrations from Claringbold et al.’s JWST analysis (REFERENCE) and the dedicated prebiosignature study now underway for LIFE (In Prep REFERENCE)

%\subsubsection{Spatial and temporal effects on retrieving planet parameters}

LIFE may be able to detect variations in surface temperature, equilibrium temperature, and Bond albedo \citep{mettler2024}, critical for rocky planets around ultracool stars, which are predicted to exhibit stronger spatial and temporal variations in physical and chemical properties compared to Earth \cite[e.g.][]{braam2025}. Combining comprehensive 4D climate–chemistry modelling with synthetic mid-infrared observations, \citet{braam2026} show that LIFE can characterise spatial and temporal variations in the atmospheres of nearby M-star rocky exoplanets, breaking critical degeneracies and false positives for phase-dependent features such as ozone.

\vspace{-10pt}
\subsection{Cold Giant Planets}

In addition to habitable planets, LIFE will be able to detect Jupiter and Neptune-sized planets orbiting up to a few AU away from their host stars within 20pc \citep{alei2024}. \texttt{LIFEsim} modeling shows that dozens of known cold giants are accessible to LIFE in modest integration times (10–100~hr) \citep{carrion2023}. Cold giants are less likely to transit and are faint in reflected light. In the mid-infrared (MIR), however, these planets emit detectable thermal radiation. LIFE offers the contrast ($\sim 10^{-7}$) and angular resolution ($\sim 50$ mas) to detect their emission \citep{alei2024}. LIFE can measure their emission spectra and break degeneracies on mass, radius, and albedo. By surveying a broader stellar sample, LIFE is also expected to detect new long-period giants that lie beyond the reach of current detection techniques.

%Cold Jupiter and Neptune-sized planets ($T_{\mathrm{eff}} \lesssim 300$ K) are common outcomes of planet formation \citep{Mayor2011}, but remain poorly characterized. These objects hold clues to the formation mechanisms and the dynamical evolution of planetary systems. Their chemical compositions, especially the metallicity and the C/O ratio, are sensitive to the location of formation and the balance between solid and gas accretion \citep{Oberg2011, Madhusudhan2014}. Cold giants also regulate the architecture of the planet's system and the delivery of volatiles to inner planets \citep{Morbidelli2012}.

%Cold giants are typically non-transiting and faint in reflected light, making them inaccessible to transit and optical coronagraphy. In the mid-infrared (MIR), however, these planets emit detectable thermal radiation. A mission like LIFE, using nulling interferometry in the 5--18 µm range, offers the contrast ($\sim 10^{-7}$) and angular resolution ($\sim 50$ mas) to detect their emission at separations of a few AU around stars within $\sim$20~pc \citep{Alei_2024}. 

%Unlike JWST and ELTs, which are limited by inner working angle and thermal background \citep{luhman2023jwstnirspecobservationscoldestknown, BowensRubin2025}, LIFE can spatially resolve and spectrally characterize mature cold giants.

MIR spectroscopy enables retrieval of key molecules such as CH$_4$, H$_2$O, NH$_3$, CO, and PH$_3$ \citep{M_ller_2013, encrenaz2014}. These tracers constrain atmospheric metallicity and C/O ratio, as well as non-equilibrium chemistry due to vertical mixing. LIFE’s spectral range includes features sensitive to cloud decks and temperature inversions. Cold giants with high cloud opacity in the near-IR remain bright in the MIR, as seen in JWST observations of $\epsilon$~Indi~Ab \citep{Matthews_2024}. 

%LIFE's cold giant science builds on legacy concepts like TPF-I and Darwin \citep{Beichman1999, Cockell2009}, but with higher technological readiness and a known target list. 

Simultaneous detection of terrestrial planets and gas giants will enable planetary system-level studies. MIR spectra of cold giants will help constrain substellar evolution, volatile content, and cloud structure and aid comparative planetology of gas-rich worlds and inform planet formation models.
\vspace{-10pt}

\subsection{Protoplanets and exocomets}

 Magma ocean planets are extremely hot (\(\sim\)500–3,000 K) and bright in the near- to mid-infrared, making them promising LIFE targets. The magma ocean phase is a key stage in early planetary evolution, shaping interior differentiation, volatile distribution, and development of secondary atmospheres. Earth-sized magma ocean planets in habitable zones could be detected up to \(\geq\)100 pc away in minutes to hours of integration time, allowing imaging much further than the solar neighborhood \citep{Cesario_2024}. Detecting Earth-sized protoplanets in this phase could help constrain the conditions of an early Earth atmosphere. Protoplanets in M-dwarf habitable zones tend to offer the most favorable detection conditions, needing less integration time than those around solar-type stars. This trend holds for systems \(\lesssim\)70 pc away, but beyond this distance, solar-type habitable zone protoplanets have more favorable detection conditions than M-star habitable zone planets due to angular resolution limitations.

 %Unlike typical LIFE targets, detectability of magma ocean early-Earths depends more on wavelength range than on aperture size \citep{Cesario_2024}. Shorter wavelengths ($\sim$ 3-6 µm) yield higher signal to noise ratios (S/N) and therefore important to better constrain the temperatures of such planets \citep{Cesario_2024}. 
 %Using \texttt{LIFEsim},  found that  Required integration time increases with distance and decreases exponentially with object temperature.

%Although young M dwarfs are brighter early in their lives, their age has little effect on required integration time \citep{Bonati_2019}. 

%For protoplanets T \(\gtrsim\) 1500 K, 1000x solar zodiacal dust densities minimally affect detectability. On the other hand, cooler (\(\lesssim\)1000 K) protoplanets become undetectable under such conditions \citep{Cesario_2024}. This is because exozodiacal noise tend to be a more significant error at longer-wavelengths \citep{Quanz2022_LIFEI}.

%\subsection{Exocomets}

%Small solid bodies such as comets and asteroids (planetesimals) arise during planet formation and are therefore expected to be abundant especially in young systems \citep{Johansen2015, drazkowska2023planetformationtheoryera}. 

Exocomets have not been directly imaged, but many have been indirectly identified in the $\beta$ Pictoris system  \citep[see e.g.][]{Ferlet1987,refId0}. Some planetesimals are scattered close to their host stars, with disruption events or forming a coma that may be visible in the infrared. Simulations of exocomets with LIFE for the $\beta$ Pic system show that even with high levels of exozodiacal dust, much of the parameter space yielded high S/N with 10hr integration times \citep{Janson_2023}. Indeed, $\beta$ Pic might be too dense to resolve and less crowded systems such as $\epsilon$ Eridani, Fomalhaut, or AU Mic along with 30 other potential exocomet systems \citep{strom2020} may provide better detections that could constrain comet parameters and potentially the mineralogy of their dust.% A fraction of Solar System comets have strong silicate emission features, and so it may become possible to study the mineralogy of individual exocometary bodies as those features could be detectable with LIFE.

 \vspace{-10pt}

\subsection{AGNs, galactic structures, circumstellar disks and star forming regions}
 
Studies of the related DARWIN/TPF-I missions suggest that LIFE's mission requirements could make it valuable for imaging star forming regions, the cores of active galactic nuclei (AGNs), the inner regions of dusty circumstellar disks, and complex galactic structures. The 4-18.5µm wavelength range probed by LIFE is well-suited for determining chemical properties of AGNs, and its high angular resolution enables detection of AGN jets \citep{TPF-I_SWG_2008}. Furthermore, prior simulations suggest that LIFE could resolve high-redshift galactic structures, OB associations, and even microlensing events.%, albeit over a narrower wavelength range than JWST.
\vspace{-10pt}

\section{Synergies with NASA's Habitable Worlds Observatory}

After recommendation from the National Academy of Sciences in the Astro 2020 Decadal Survey, NASA is pursuing the Habitable Worlds Observatory (HWO), a 6-8 meter UV, visible, and near-infrared telescope as their next flagship mission \citep{feinberg2024}. As the development of both HWO and LIFE progress, it is useful to quantify potential synergies between the two telescopes as well as to explore how UK science and technology will benefit from an investment in both. 

\begin{figure}[h!]
%\begin{wrapfigure}{L}{0.7\textwidth}
%\vspace{-13pt} % tighten space above
\centering
    \includegraphics[width=1.01\textwidth ]{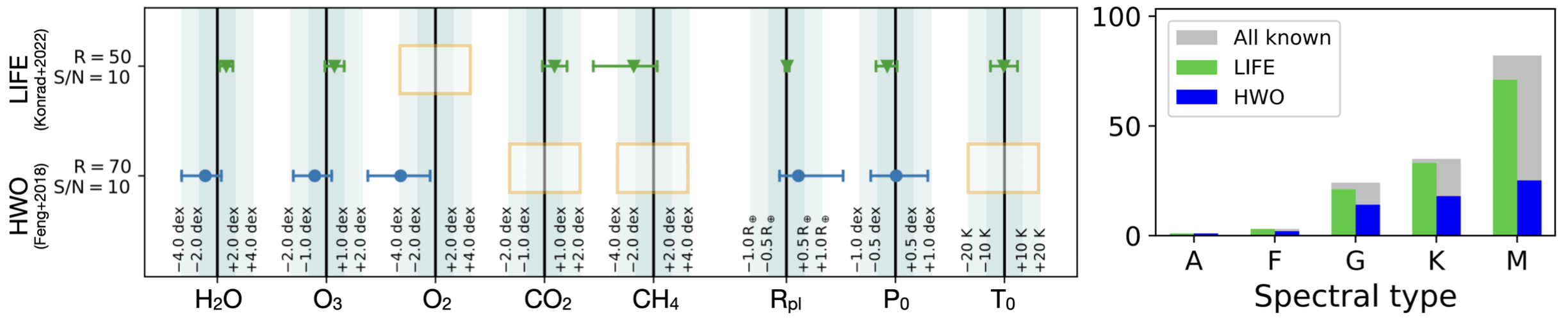} \caption{Fig. 3: Retrieval results of molecular abundances and planetary parameters for a simulated thermal emission spectrum for an Earth-twin with LIFE and for HWO (left) \citep[figure adapted from][]{konrad2022} and expected planet yields with spectral type (right) \citep[adapted from][]{carrion2023}}  \label{hwovslife}
\end{figure}

HWO and LIFE will cover a complementary parameter space in terms of dimensions, stellar class, and semi-major axis \citep{carrion2023}. While HWO will focus on FGK stars, whose habitable zone is further out beyond the inner working angle limit of the coronagraph, LIFE will explore FGK as well as M targets, probing 30 - 50 planets closer to their stars (see right panel of Fig. \ref{hwovslife}). HWO will obtain spectra of at least 25 Earth analogues around FGK stars, achieving contrast levels of 10$^{-10}$ through high-contrast coronagraphy. Various channels still in discussion are expected to cover the UV to near-IR (0.3 - 1.7 µm, tbc). HWO will be able to characterize the UV of the host star, important to understanding the planet-star connection \citep[see e.g.][]{rugheimer2013,rugheimer2015b}.

%Various channels still in discussion are expected to be in the UV (0.3-0.5 µm) at R=7, in the optical (0.5-1 µm) at R=140, and in the near-infrared (1-1.85 µm) at R$\geq$70.

%\vspace{-10pt} % tighten space above
%\end{wrapfigure}

Currently there are $\sim$ 50 known exoplanets within 20pc that could be observed by both LIFE and HWO \citep{carrion2023}. For these targets we could acquire both the reflected and the emitted spectrum. \citet{alei2024} explore this unprecedented characterisation potential for a cloud-free Earth-twin. Using simulated atmospheric retrievals of reflected-light spectra, they find HWO can correctly determine the bulk components of the atmosphere (N$_2$ and O$_2$) from Rayleigh scattering and sharp O$_2$ lines in the visible range. Ozone is detectable through a deep UV band, which HWO might be able to characterise at a very coarse resolution (R=7). Water vapour, ocean glint, and surface biosignatures may also be detectable in the VIR/NIR region under very favorable conditions \citep{feng2018, damiano2022, cowan2025, Ulses_2025, krissansen2025}. 

The mid-IR emission spectra between 4 and 18.5 µm at R=100 could assess the thermal structure of the atmosphere, as well as critical habitability indicators as CO$_2$, H$_2$O, and the biosignatures O$_3$ and CH$_4$ \citep{konrad2022,konrad2024}. Of note, HWO is not able to get a constraint on the surface temperature whereas LIFE can. As well, LIFE is better able to constrain the surface pressure and radius which are vital  planetary context indicators to assess habitability. CO$_2$ has weak features in the NIR and is not accessible to HWO at Earth concentrations, but is accessible to LIFE at lower concentrations. Additionally, CH$_4$ remains challenging in reflected light due to the weak feature and low flux in the near-IR but it has a strong feature in the mid-IR.

%LIFE will not observe the O$_2$ A band, which is valuable for retrieving an accurate abundance of oxygen as the IR feature of O$_3$ is less sensitive to abundance. HWO will also be able to get the 
See Fig. \ref{hwovslife} for an overview of which atmospheric molecules and planetary parameters are best constrained for an Earth-twin with each instrument. Using both HWO and LIFE will allow us to retrieve all the main components of the atmosphere, as well as cloud and surface albedo information (from HWO), and temperature-pressure and radius estimates (from LIFE). Combining the two wavelength regimes will open two separate but complementary windows into the atmosphere of an exoplanet (left panel of Fig. \ref{hwovslife}), providing crucial information on its climate and habitability.% More studies will be necessary to explore HWO-LIFE synergies across different atmospheric scenarios, stellar classes, and instrument architectures as they become finalised.
\vspace{-10pt}
\section{Strategic Context}

The ESA Senior Committee Report\footnote{\url{https://www.cosmos.esa.int/documents/1866264/1866292/Voyage2050-Senior-Committee-report-public.pdf}} prioritised a mid-IR mission for L5: ``Therefore, launching a Large mission enabling the characterisation of the atmosphere of temperate exoplanets in the mid-infrared should be a top priority for ESA within the Voyage 2050 timeframe.'' In addition, there is widespread support for the LIFE mission in Europe and globally. At the November 2025 LIFE meeting in Barcelona there were 250 scientists ranging from exoplanet modeling to instrument science and space engineering from 32 countries. As well, the general public is extremely interested in life detection.

A leading role in LIFE will benefit UK industry by continuing our leadership in space engineering and infrared instrumentation, and growing the UK space economy with academic-industrial partnerships. Partnering with LIFE represents an opportunity to build upon the scientific and technical expertise created by the UK leadership in prior missions such as JWST, Plato, and Ariel. With LIFE we will strengthen European competitiveness in one of the most vibrant areas of modern astrophysics and contribute to the exoplanet space mission designed to observe an oxidizing + reducing gas biosignature combination \citep{alei2024}, the most robust detection for life as we know it.

\vspace{-10pt}

\section{Proposed Approach}

UKSA’s early modest investment has led to larger contracts being awarded as in the case with the UK contribution to the Solar Orbiter leading to the Astrium UK contract. In the same vein, we propose a feasibility study to look at a potential partnership with the UK Astronomy Technology Centre (UK ATC) which built and tested MIRI, the mid-infrared Instrument for JWST and is building METIS for the ELT. The co-PI of LIFE and lead of the instrument team, Adrian Glauser, also has historical ties to the UK ATC when he was the Swiss National Project Manager for JWST-MIRI. In the short term, a small pilot study to identify the requirements for building the infrared spectrometer for LIFE would cement the UK taking a leadership role in being a national partner of this future mission.

%\footnote{\url{https://www.esa.int/Science_Exploration/Space_Science/ESA_contracts_Astrium_UK_to_build_Solar_Orbiter}}

\vspace{-10pt}

\section{Proposed technical solution and required development}
Directly imaging temperate, rocky exoplanets in the mid-IR requires planet-star contrast ratios of $\sim\nobreak10^{-7}$. This maps to an average null depth of at least 10$^{-5}$ for LIFE which will use nulling interferometry to achieve this requirement \citep{glauser2024}. This involves combining the light from 4 formation flying collector spacecraft into a 5th combiner spacecraft with a $\pi$ phase shift between beam pairs to create destructive interference of the on-axis stellar signal. Rotating the system then allows the exoplanet signal to be modulated. This technique has the direct advantage over a coronagraph in that it can resolve planetary signals that would typically be contained within the stellar point spread function, but comes with three main technological challenges that need to be tackled before launch. 
	
First, formation flying precision is the main technological requirement. ESA’s Proba-3 mission recently achieved sub-millimetre precision using optical metrology \citep{penin2020}, building on LISA Pathfinder’s 10 pm/$\sqrt{\text{Hz}}$ noise performance \citep{armano2018}. Precursors like SILVIA are explicitly designed to bridge this gap for interferometry missions like LIFE \citep{ito2025} along with developments in deformable mirrors for optical path length correction at NOVA in the Netherlands.

Second, photonic beam combination is being developed on the ground. VLTI instruments like Asgard and NOTT target contrasts of 10$^{-5}$ \citep{garreau2024}, while GLINT at Subaru has already demonstrated $2.5 \times 10^{-4}$ on sky \citep{norris2020}. Photon-counting mid-IR detectors are being developed at SRON. In parallel, SiGe waveguide chips are under development to provide highly compact and stable mid-infrared components \citep{fedeli2018}.

Third, cryogenic optical control and mid-IR detectors are under development. ETH Zürich’s Nulling Interferometer Cryogenic Experiment (NICE) has achieved stable nulls of $7\times10^{-6}$ at room temperatures. The next milestone is 15 K, aimed at validating the $10^{-5}$ requirement. This is complemented by advances in mid-IR Microwave Kinetic Inductance Detectors (KIDs), offering promising single-photon sensitivity in the mid-IR \citep{ulbricht2021}.

\vspace{-10pt}

\section{UK Leadership and Capability}

The UK exoplanet community is very active with faculty and groups at the following universities: Imperial College, Keele, Kings College London, Queen Mary University of London, Queens University Belfast, the OU, UCL, Birmingham, Bristol, Cambridge, Cardiff, Lancashire, Edinburgh, Exeter, Leeds, Leicester, Hertfordshire, Manchester, Oxford, St Andrews, and Warwick. Yearly the UKExoM meeting attracts 150-200 participants from around the UK to discuss the latest exoplanet research in the country.

The UK Space Agency commissioned an Ariel interim report showing that UK authors contribute 12-16\% of all exoplanet papers \citep{knowspace2025}. This well-established community will be well-served by investment in the development of the Large Interferometer for Exoplanets (LIFE). A UKSA-funded feasibility study is an important first step. If successful, this will help cement a continued strategic engagement and leadership with the LIFE mission.
\vspace{-10pt}

\subsection{UK Astronomy Technology Centre (UK ATC)}

The UK ATC skill set includes expertise in several technologies which will be critical to the success of the LIFE mission.  We have identified two particularly promising areas where a relatively low investment in effort can place us at the heart of what will be a long and prestigious project. 
 
First, our heritage in building spectrometers for the major ground based observatories (e.g. METIS on the ELT) and MIRI for JWST, leads us to possibly contribute to the spectrometer development, where a small solid state optical (i.e non-photonic) spectrometer would be coupled to a state of the art kinetic induction detector (KID) provided by SRON in the Netherlands, and installed in the LIFE interferometer test bed at ETH in Zurich. 
 
Second, the ATC’s adaptive optics group is well suited to contributing to the system design and modelling of the tip/tilt and deformable mirror optics which form part of LIFE’s beam collection optics.

\vspace{-10pt}

\section{Partnership Opportunities}

Involvement with the LIFE mission will position the UK and Europe to lead this next generation space telescope. LIFE is a European mission, with the current headquarters at ETH Zürich. The LIFE Science community is global with active projects testing the LIFE mission requirements, instrument simulators, and technology development in Europe, Asia, North \& South America and Australia. 

The UKATC has expertise in infrared detectors and their prior expertise in contributing to MIRI on JWST and their planned role in developing for HWO makes the UK well positioned to provide technology and instrument development for LIFE along with other member countries.

ESA's Voyage 2050 report has already highlighted that the main priority for the next L5 mission should be a LIFE like mission. Private industry has also expressed interest in joining space exploration and research. The UK's involvement now will allow the UK to position itself as a leader and partner in one of the most ambitious space missions of our generation. NASA is leading HWO. Europe will lead LIFE. The UK should cement our place in being a strategic partner in both and lead the development an infrared spectrometer and contribute to the system design and deformable mirror optics for LIFE.

%Swiss National Science Foundation, ETH Zurich SPACE

\newpage
\noindent\large{\textbf{Co-authors with affiliations:}}
\vspace{2mm}
\normalsize

\noindent
Sarah Rugheimer - University of Edinburgh \\
Aiden Weatherbee - York University, Canada \\
James Fecanin - University of Edinburgh \\
Alistair Glasse - UK Astronomy Technology Centre \\
Paul Rimmer - University of Cambridge \\ 
Eleonora Alei - NASA Goddard Space Flight Center \\ 
Esther Wang - University of Bern \\ 
Marrick Braam - University of Bern \\ 
Sascha P. Quanz - ETH Zürich \\
Adrian Glauser - ETH Zürich \\ 
Alexander Archibald - University of Cambridge \\ 
Beth Biller - University of Edinburgh \\
Mark Booth - UK Astronomy Technology Centre \\ 
Tereza Constantinou - University of Cambridge \\ 
Gregory Cooke - University of Cambridge \\ 
Daniel Dicken - UK Astronomy Technology Centre \\
Trent Dupuy -  University of Edinburgh \\
Mei Ting Mak - University of Oxford \\ 
Paul Palmer - University of Edinburgh \\
Tim Pearce - University of Warwick \\
Vito Squicciarini - University of Exeter \\
Amaury Triaud - University of Birmingham \\
Floris van der Tak - Space Research Organization Netherlands \\
Sergey Yurchenko - University College London \\

\noindent\large{\textbf{UK co-signatories:}}

\vspace{2mm}
\normalsize
\noindent 
Suzanne Aigrain - University of Oxford \\
David Armstrong - University of Warwick \\
Joanna Barstow - Open University \\
Martin Barstow - University of Leicester \\
Matthew Battley - Queen Mary University of London \\
Jayne Birkby - University of Oxford \\
Amy Bonsor - University of Cambridge \\
Richard Booth - University of Leeds \\
Ryan Boukrouche - University of Southampton \\
David J. A. Brown - University of Warwick \\
Heather Cegla - University of Warwick \\
Xueqing Chen - University of Edinburgh \\
Katy L. Chubb - University of Bristol \\
Alastair Claringbold - University of Warwick \\
Charles Cockell - University of Edinburgh \\
Maureen Cohen - The Open University \\
Matthew Cole - University of Edinburgh \\
Andrzej Fludra - UKRI STFC, RAL Space \\
Clémence Fontanive - University of Edinburgh \\
Gergely Friss - University of Edinburgh \\
Vincent Geers - UK Astronomy Technology Centre, UKRI STFC \\
Edward Gillen - Queen Mary University of London \\
Helen Grant - University of Kent \\
Claire Guimond - University of Oxford \\
Carole Haswell - The Open University \\
Éric Hébrard - University of Exeter \\
Sasha Hinkley - University of Exeter \\
Aiza Kenzhebekova - University of Edinburgh \\
James Kirk - Imperial College London \\
Thaddeus Komacek - University of Oxford \\
Adam Koval - University of Edinburgh \\
Yi Lu - ESO/University of Exeter \\
Ryan MacDonald - University of St Andrews \\
Nikku Madhusudhan - University of Cambridge \\
Sean McMahon - University of Edinburgh \\
Nathan Mayne - The University of Exeter \\
Annelies Mortier - University of Birmingham \\
Harrison Nicholls - University of Oxford \\
Cyrielle Opitom - University of Edinburgh \\
Vatsal Panwar - University of Birmingham \\
Mia Belle Parkinson - University of Edinburgh \\
Anjali Piette - University of Birmingham \\
Raymond Pierrehumbert - University of Oxford \\
John Maurice Campbell Plane - University of Leeds \\
Don Pollacco - University of Warwick \\
Ken Rice - Institute for Astronomy, University of Edinburgh \\
Richard Robinson - University of Portsmouth \\
David Rosario - Newcastle University \\
Felix Sainsbury-Martinez - University of Leeds \\
Subhajit Sarkar - Cardiff University \\
Denis Sergeev - University of Bristol \\
Paul Streeter - Open University \\
Ben Sutlieff - University of Edinburgh \\
Matthew Iain Swayne - University of Glasgow \\
Jake Taylor - Department of Physics, University of Oxford \\
Jonathan Tennyson - University College London \\
Samantha J. Thompson - University of Cambridge \\
Eleni Tsiakaliari - Open University \\
Daniel Valentine - University of Bristol \\
Dimitri Veras -  University of Warwick \\
Hannah Wakeford - University of Bristol \\
Catherine Walsh - University of Leeds \\
Peter J. Wheatley - University of Warwick \\
Mark Wyatt - University of Cambridge \\

\noindent\large{\textbf{International co-signatories:}}
\vspace{2mm}
\normalsize

\noindent 
Jason Dittmann - University of Florida \\
Jonathan Fortney - University of California, Santa Cruz \\
Lisa Kaltenegger - Cornell University \\
Daniel Kitzmann - University of Bern \\
Tim Lichtenberg - University of Groningen \\
Victoria Meadows - University of Washington/SETI Institute \\
Adam Muzzin - York University \\
Heike Rauer - Freie Universität Berlin; German Aerospace Center \\
Dimitar Sasselov - Harvard University \\
Sara Seager - MIT \\
Benjamin Taysum - German Aerospace Center (DLR) \\

%\normalsize

\bibliographystyle{aasjournal}
%\bibliographystyle{mnras}

%\bibliography{bib}

\clearpage

\end{document}